# Non-equilibrium Green's function treatment of phonon scattering in carbon nanotube transistors

Siyuranga O. Koswatta,[*] Sayed Hasan, and Mark S. Lundstrom

School of Electrical and Computer Engineering, Purdue University, West Lafayette, Indiana 47907, USA

M. P. Anantram

Department of Electrical and Computer Engineering, University of Waterloo, Waterloo, Canada

Dmitri E. Nikonov

Technology and Manufacturing Group, Intel Corp., SC1-05, Santa Clara, California 95052, USA

*Abstract* - We present the detailed treatment of dissipative quantum transport in carbon nanotube field-effect transistors (CNTFETs) using the non-equilibrium Green's function formalism. The effect of phonon scattering on the device characteristics of CNTFETs is explored using extensive numerical simulation. Both intra-valley and inter-valley scattering mediated by acoustic (AP), optical (OP), and radial breathing mode (RBM) phonons are treated. Realistic phonon dispersion calculations are performed using force-constant methods, and electron-phonon coupling is determined through microscopic theory. Specific simulation results are presented for (16,0), (19,0), and (22,0) zigzag CNTFETs that are in the experimentally useful diameter range. We find that the effect of phonon scattering on device performance has a distinct bias dependence. Up to moderate gate biases the influence of high-energy OP scattering is suppressed, and the device current is reduced due to elastic back-scattering by AP and low-energy RBM phonons. At large gate biases the current degradation is mainly due to high-energy OP scattering. The influence of both AP and high-energy OP scattering is reduced for larger diameter tubes. The effect of RBM mode, however, is nearly independent of the diameter for the tubes studied here.

[*] Email address: koswatta@purdue.edu

## I. INTRODUCTION

Since the first demonstration of carbon nanotube (CNT) field-effect transistors in 1998 [1,2], there has been tremendous progress in their performance and the physical understanding [3]. Both electronic as well as optoelectronic devices based on CNTs have been realized, and the fabrication processes have been optimized. Ballistic transport in CNTs has been experimentally demonstrated for low-bias conditions at low temperatures [4,5]. High-performance CNT transistors operating close to the ballistic limit have also been reported [6,7,8]. The experimentally obtained carrier mobilities are of the orders $10^4 \sim 10^5$ cm$^2$/Vs [9,10] so exceptional device characteristics can indeed be expected. Current transport in long metallic CNTs, however, is found to saturate at ~ 25 µA at high biases, and the saturation mechanism is attributed to phonon scattering [11]. On the other hand, for short length metallic tubes, the current is found not to saturate but to increase well beyond the above limit [12,13].
Nevertheless, carrier transport in these shorter tubes is still influenced by phonon scattering, and warrants a detailed physical understating of the scattering mechanisms due to its implications on device characteristics for both metallic as well as semiconducting CNTs.

There have been many theoretical studies on the calculation of carrier scattering rates and mobilities in CNTs using semiclassical transport simulation based on the Boltzmann equation [14,15,16,17,18,19,20]. Similarly, phonon mode calculations for CNTs are also performed with varying degrees of complexity: continuum and force-constant models [21,22,23] to first-principles based methods [24,25,26]. The determination of electron-phonon (e-ph) coupling strength is performed using tight-binding calculations [27,28,29] as well as first-principles techniques [30]. It has been shown, however, that the influence of phonon scattering on device performance depends not only on the phonon modes and e-ph coupling, but also on the device geometry [31,32]. Therefore, in order to ascertain the impact of phonon scattering on the device performance, aforementioned calculations should be done in the context of specific device geometry. To that end, phonon scattering in CNT transistors has been treated using the semiclassical Boltzmann transport to determine its effects on device characteristics [31,33]. Semiclassical transport, however, can fail to rigorously treat important quantum mechanical effects, such as band-to-band tunneling, that have been deemed important in these devices [34,35,36]. Therefore, a device simulator based on dissipative quantum transport that rigorously treats the effects of phonon scattering will be essential for the proper assessment of CNT transistor characteristics, and to gain a deeper understanding of carrier transport at nanoscale.

The non-equilibrium Green's function (NEGF) formalism has been employed to describe dissipative quantum transport in nanoscale devices [37,38,39]. It has been used to treat the effects of phonon scattering in CNT Schottky barrier transistors (SBFETs) [40,41]. It has also been successfully used to investigate the impact of phonon scattering, and to explore interesting transport mechanisms such as phonon-assisted inelastic tunneling, in CNT metal-oxide semiconductor field-effect transistors (MOSFETs) with doped source and drain contacts (hereafter, simply referred to as CNTFETs) [32,34,35,36]. The NEGF simulation of ballistic transport in CNTFETs is reported in [42]. Here, we extend the previous work [42], and present the detailed simulation technique employed for the treatment of phonon scattering in them. Section II describes the tight-binding scheme, the self-consistent electrostatics, and the treatment of e-ph coupling for NEGF modeling of the CNTFETs. Section III summarizes the numerical procedures used for the simulation of phonon scattering in the self-consistent Born approximation. Section IV, followed by the conclusion in section V, has the detailed simulation results, and discusses the impact of phonon scattering on CNTFET characteristics. It compares the diameter dependence of the effect of phonon scattering in (16,0), (19,0), and (22,0) zigzag CNTs (i.e.: *mod(n-m,3)* = 1 type) that are in the experimentally useful diameter range (1.2nm ~ 1.8nm), below which the contact properties degrade, and above which the bandgap is too small for useful operation [43].

## II. METHOD

### A. Treatment of Transport by NEGF

A detailed description of the NEGF modeling of ballistic transport in CNTFETs is described in [42]. Here we present a brief overview of that device model for the sake of completeness. The device Hamiltonian used in this study is based on the atomistic nearest-neighbor $p_z$-orbital tight-binding approximation [21]. The device geometry, shown in Fig. 1(a), is a CNT MOSFET with doped source and drain regions ($L_{SD}$) and a cylindrical wrap-around metallic gate electrode over the intrinsic channel region ($L_{ch}$). The gate oxide with thickness $t_{OX}$ covers the full length of the tube. We employ artificial heavily doped extension regions, $L_{ext}$. They do not influence the transport in the working part of the transistor, but useful for better numerical convergence purposes when phonon scattering is present (however are not necessary for ballistic simulations). The cylindrical geometry of this device ensures symmetry in the angular direction thus drastically simplifying the mode-space treatment of electron transport [42,44]. It also permits the treatment of self-consistent electrostatics using 2D finite difference method [42]. The source and drain electrodes are treated as quasi-continuum reservoirs in thermal equilibrium and are modeled by the contact self-energy functions as in [42].

The NEGF model of the CNTFET used for transport simulations is shown in Fig. 1(b). Here, $H_{pz}$ is the device Hamiltonian and the self-energies $\Sigma_{S/D}$ represent the semi-infinite ideal source/drain contacts. $\Sigma_{scat}$ is the self-energy for e-ph interaction, and one sets $\Sigma_{scat} = 0$ for the ballistic approximation. A detailed specification of $\Sigma_{scat}$ is presented later in section II.D. Finally, the retarded Green's function for the device in the matrix form is given by [37],

$$G(E) = \left[ \left( E + i\eta^+ \right) I - H_{pz} - \Sigma(E) \right]^{-1} \quad (1)$$

where $\eta^+$ is an infinitesimal positive value, and $I$ the identity matrix [37].
The self energy contains the contributions from all mechanisms of relaxation; the source and drain electrodes, and from scattering [37]

$$\Sigma(E) = \Sigma_S(E) + \Sigma_D(E) + \Sigma_{scat}(E) \quad (2)$$

Note that in Eq. (2) the self-energy functions are, in general, energy dependent.

In the mode-space treatment of an (*n*,0) zigzag CNT, the dependence of the electronic state on the angle along the tube's circumference, $\varphi$, is expanded in a set of circular harmonics exp(*im*$\varphi$) with the angular quantum number, *m*. It spans the integer values of 1 to $2n$, or, equivalently, $-n+1$ to $n$. Integer values on $m$ outside this range would produce equivalent harmonics at the crystal lattice sites. The total Hamiltonian splits into independent matrices for subbands associated with each value of $m$ [42], giving rise to a 1D Hamiltonian with two-site unit cell, as schematically shown in Fig. 1(c), where each site corresponds to one of two non-equivalent real-space carbon rings, A or B. The period of the zigzag tube in the longitudinal direction contains 4 such rings, ABAB, and has length $3a_{cc}$ [21], where $a_{cc} = 0.142 nm$ is the carbon-carbon bond length in graphene. Therefore the average distance between rings is

$$\Delta z = \frac{3a_{cc}}{4}. \tag{3}$$

The diameter of the zigzag nanotube is [21]

$$d_t = \frac{n\sqrt{3}a_{cc}}{\pi} \tag{4}$$

The mode-space transformation procedure of the real-space atomistic tight-binding Hamiltonian is well described in [42], and is not repeated here. The two-site unit cell, as expected, gives rise to two subbands corresponding to the conduction and the valence band. The Hamiltonian matrix for the subbands with angular quantum number $m$ in an $(n,0)$ zigzag CNT is then given by [42],

$$H_{pz} = \begin{bmatrix} U_1 & b_{2m} & & & & \\ b_{2m} & U_2 & t & & 0 & \\ & t & U_3 & b_{2m} & & \\ & & & \ddots & & \\ & 0 & & t & U_{N-1} & b_{2m} \\ & & & & b_{2m} & U_N \end{bmatrix}_{N \times N} \tag{5}$$

where $b_{2m} = 2t\cos(\pi m/n)$, $t \approx 3 eV$ is the nearest neighbor hopping parameter, and $N$ is the total number of carbon rings along the device. Here, the diagonal elements $U_j$ correspond to the on-site electrostatic potential along the tube surface. All electronic subbands in a CNT are four-fold degenerate: due to two spin states and the valley degeneracy of two [21]. The valley degeneracy comes from the two subbands with the same energy dispersion, but different *m*-values. Each subband can be represented as a cut of the graphene 2D Brillouin zone by a line with a constant momentum $k_y$. In this paper we equate momentum with wavevector, having the dimension of inverse length. The cuts closest to the K-points of graphene correspond to lowest-energy conduction subbands as well as highest-energy valence subbands, and correspond in zigzag tubes to angular momenta $m_{L1} = round(2n/3)$ and $m_{L2} = round(4n/3)$.

Level broadening is defined as follows and can be shown [37] to be

$$\Gamma(E) \equiv i\left[\Sigma(E) - \Sigma^\dagger(E)\right] = \Sigma^{in}(E) + \Sigma^{out}(E), \tag{6}$$

where $\Sigma^\dagger$ represents the Hermitean conjugate of $\Sigma$ matrix defined by Eq. (2). Here, $\Sigma_{scat}^{in/out}$ are the in/out-scattering functions (see below). The same relations apply separately to each mechanism of relaxation.

For a layered structure like the carbon nanotube, the source self-energy function $\Sigma_{source}$ has all its entries zero except for the (1,1) element. That is [42],

$$\Sigma_S(i \neq 1, j \neq 1) = 0 \tag{7}$$

and,

$$\Sigma_S(1,1) = \alpha_{source} - \sqrt{\alpha_{source}^2 - t^2}, \quad \alpha_{source} = \frac{(E - U_1)^2 + t^2 - b_{2m}^2}{2(E - U_1)} \tag{8}$$

Similarly, $\Sigma_D$ has only its (*N,N*) element non-zero and it is given by equations similar to (7) and (8) with $U_1$ replaced by $U_N$. As mentioned earlier, $\Sigma_{S/D}$ self-energies rigorously capture the effect of semi-infinite contacts on the device. With this we can define the in- and out-scattering functions for contact coupling,

$$\Sigma^{in}_{S/D}(E) = \Gamma_{S/D}(E) f(E - E^F_{S/D}) \qquad (9)$$

$$\Sigma^{out}_{S/D}(E) = \Gamma_{S/D}(E)\left[1 - f(E - E^F_{S/D})\right] \qquad (10)$$

where $f(E)$ is the Fermi distribution, and $E^F_{S/D}$ are the source and drain Fermi energies, respectively. The in/out-scattering functions for e-ph interaction are discussed later in section II.D. The electron and hole correlation functions are then given by,

$$G^n(E) = G\Sigma^{in}G^\dagger \qquad (11)$$

$$G^p(E) = G\Sigma^{out}G^\dagger \qquad (12)$$

where the energy dependence of the Green's function and in/out-scattering functions is suppressed for clarity. The spectral function is [37]

$$A(E) \equiv i\left(G(E) - G^\dagger(E)\right) = G^n(E) + G^p(E) \qquad (13)$$

Note that the electron and hole correlation functions, $G^{n/p}_{i,j}(E,m)$, are matrices defined in the basis set of ring numbers $i,j$ and subbands $m$ (we will imply the last index in the rest of the paper). Thus the diagonal elements, $G^{n/p}_{j,j}(E,m)$, correspond to the energy density of carrier occupation at those basis sites (single carbon ring, A or B, in a specific subband) with a given energy $E$. So the total electron/hole density (per unit length) at a site $z_j$ is given by,

$$n(z_j) = \sum_{m,s} \frac{1}{\Delta z} \int_{-\infty}^{+\infty} \frac{G^n_{j,j}(E,m)}{2\pi} dE \qquad (14)$$

$$p(z_j) = \sum_{m,s} \frac{1}{\Delta z} \int_{-\infty}^{+\infty} \frac{G^p_{j,j}(E)}{2\pi} dE \qquad (15)$$

where summation is performed over the spin and subband variables, and produces the degeneracy factor of 4 (for each non-equivalent subband). In the view of Eq. (13) one recognizes that the spectral function is proportional to the density of states which is traditionally defined [45] to include the spin summation, but is taken separately for each subband

$$g_{1D}(E, z_j) = \frac{A_{j,j}(E, m)}{\pi \Delta z} \tag{16}$$

Finally, the current flow from site $z_j$ to $z_{j+1}$ in the nearest-neighbor tight-binding scheme can be determined from [38,39],

$$I_{j \to j+1} = \sum_{m,s} \frac{ie}{\hbar} \int_{-\infty}^{+\infty} \frac{dE}{2\pi} \left[ H_{j,j+1}(m) G^n_{j+1,j}(E, m) - H_{j+1,j}(m) G^n_{j,j+1}(E, m) \right] \tag{17}$$

wherein the non-diagonal terms of the Hamiltonian Eq. (5) contain only contributions of hopping. The above equation is a general relationship, in that it is valid even under dissipative transport. Under ballistic conditions, however, Eq. (17) further simplifies (for each non-equivalent subband) to,

$$I = \frac{4e}{\hbar} \int_{-\infty}^{+\infty} \frac{dE}{2\pi} T(E) \left[ f(E - E^F_S) - f(E - E^F_D) \right] \tag{18}$$

with the transmission coefficient, $T(E)$, given by

$$T(E) = Trace \left[ \Gamma_S(E) G^r(E) \Gamma_D(E) G^{r\dagger}(E) \right] \tag{19}$$

Eq. (19) is the famous Landauer equation widely used in mesoscopic transport [37].

One can better understand the bandstructure of carbon nanotubes in by solving for the eigenvalues of the Hamiltonian (5) for zero external potential, and thereby obtaining [42] the energy dispersion relations, $E(k_z)$, versus the momentum along the length of the tube, for each subband. For the lowest conduction and the highest valence subbands, close to the K-points the graphene band edge is approximately conic, thus

$$\left(\frac{2E}{E_g}\right)^2 = 1 + \left(\frac{k_z}{\Delta k}\right)^2 \tag{20}$$

with the bandgap

$$E_g = 2v_F \hbar \Delta k \tag{21}$$

and the distance to the K-point of

$$\Delta k = \frac{2}{3d_t} \tag{22}$$

The velocity of carriers in the band is

$$v = \frac{dE}{\hbar dk_z} \tag{23}$$

Far enough from the band edge, the velocity tends to the constant value

$$v_F = \frac{3a_{cc}t}{2\hbar} \approx 10^6 \, m/s. \tag{24}$$

The 1D density of states including spin summation but only one subband (valley) can thus be expressed as

$$g_{1D}(E) = \frac{2}{\pi \hbar v(E)}. \tag{25}$$

or, in other terms

$$g_{1D}(E) = \frac{2}{\pi \hbar v_F} \cdot \frac{|E|}{\sqrt{E^2 - (E_g/2)^2}}. \tag{26}$$

*B. Poisson's Equation*

This section summarizes the implementation of self-consistent electrostatics in our simulation. The diagonal entries of the Hamiltonian in Eq. (5) contain the electrostatic potential on the tube surface, which thereby enters the NEGF calculation of charge distribution in Eqs. (14) and (15). On the other hand, the electrostatic potential and the charge distribution are coupled through the Poisson's equation as well, leading to the Poisson-NEGF self-consistency requirement shown in Fig. 3. The 2D Poisson equation for the cylindrical transistor geometry in Fig. 1(a) is,

$$\nabla^2 U(r,z) = -\frac{\rho(r,z)}{\varepsilon}. \tag{27}$$

Here, $\rho(r,z)$ is the net charge density distribution which includes dopant density as well. At this point, it should be noted that even though Eqs. (14) and (15) give the total carrier densities distributed throughout the whole energy range, what we really need for determining the self-consistent potential on the tube surface, $U_j \equiv U(r = r_{CNT}, z_j)$, is the induced charge density ($r_{CNT}$ = CNT radius). This can be determined by performing the integrals in Eqs. (14) and (15) in a limited energy range defined with respect to the local charge neutrality energy, $E_N$ [42,46]. In a semiconducting CNT, due to the symmetry of the conduction and valence bands, $E_N$ is expected to be at the mid-gap energy. Finally, the induced charge density at site $z_j$ can be calculated from [42],

$$Q_{ind}(z_j) = \frac{4}{\Delta z}\left[(-e)\int_{E_N(j)}^{+\infty}\frac{G_{j,j}^n(E)}{2\pi}dE + (+e)\int_{-\infty}^{E_N(j)}\frac{G_{j,j}^p(E)}{2\pi}dE\right] \tag{28}$$

where the first and second terms correspond to the induced electron and hole densities, respectively, with charge of the electron $e$.

Knowing the induced charge $Q_{ind}$, the net charge distribution $\rho(r,z)$ is given by,

$$\rho(r = r_{CNT}, z_j) = Q_{ind}(z_j) + N_D^+ - N_A^- \tag{29}$$

$$\rho(r \neq r_{CNT}, z) = 0 \tag{30}$$

where, $N_D^+$ and $N_A^-$ are ionized donor and acceptor concentrations, respectively. Here, it is assumed that the induced charge and the dopants are uniformly distributed over the CNT surface. Finally, Eq. (27) is solved to determine the self-consistent electrostatic potential $U_j$ along the tube surface. The finite difference solution scheme for the 2D Poisson equation is described in [42]. The calculated potential, $U_j^{new}$, gives rise to a modified Hamiltonian (Eq. (5)), eventually leading to the self-consistent loop between electrostatics and quantum transport (Fig. 3).

Even though the self-consistent procedure we have just outlined appears conceptually straightforward, it has poor convergence properties. Therefore, a non-linear treatment of the Poisson solution is used in practice, as explained in [38,47], in order to expedite the electrostatic convergence. The convergence criterion used in this process is to monitor the maximum change in the potential profile between consecutive iterations, i.e.: $\max(|U_j^{old} - U_j^{new}|) \leq U^{tol}$ where the tolerance value $U^{tol}$ is normally taken to be 1meV.

*C. Phonon Modes*

The parameters of the phonons are obviously determined by the structure of the nanotube lattice. The one-dimensional mass density of an (*n*,0) nanotube is,

$$\rho_{1D} = \frac{m_C n}{\Delta z}. \tag{31}$$

where $m_C$ is the mass of a carbon atom. The energy of a phonon of momentum $q$ (in the unconfined dimension) is $\hbar\omega_q$. The index of the phonon subband $l$ is implicitly combined with the momentum index here. The half-amplitude of vibration for one phonon in a tube of length $L$ is [45],

$$a_q = \sqrt{\frac{\hbar}{2\rho_{1D}L\omega_q}}. \tag{32}$$

For the reservoir in a thermal equilibrium at temperature $T$, the occupation of modes is given by the Bose-Einstein distribution

$$n_q = \left(\exp\left(\frac{\hbar\omega_q}{k_B T}\right) - 1\right)^{-1}. \tag{33}$$

As discussed earlier, the electron states in semiconducting CNTs have two-fold valley degeneracy with the lowest-energy subbands having angular quantum numbers $m_{L1}$ and $m_{L2}$. Electron-phonon scattering is governed by energy and momentum conservation rules. Thus, as shown in Fig. 2(a) electrons can be scattered within the same subband (intra-valley) where they do not change their angular momentum, and, such scattering is facilitated by zone-center phonons having zero angular momentum ($l = 0$). As shown in Fig. 2(b), it is also possible to have inter-valley scattering mediated by zone-boundary phonons having angular quantum number $l = |m_{L1}-m_{L2}|$. There can also be scattering to higher energy subbands assisted by phonon modes with $l \neq 0$ and $l \neq |m_{L1}-m_{L2}|$ [14,18], however we do not discuss results for such processes in this paper. We have performed phonon dispersion calculations using the force-constant methods described in [21,48]. As a result of this analysis, the matrix element for the electron-phonon interaction is expressed via the deformation potential, $J_1 = 6\text{eV/Å}$, and the dimensionless matrix element as follows: $|K_q| = J_1 |M_q|$. Zone-center and zone-boundary phonon dispersions for a (16,0) zigzag CNT are shown in Fig. 4(a) and 4(b), respectively. It is seen that the representation of phonon modes according to fundamental polarizations, such as longitudinal (L), transverse (T), and radial (R), can only be done for zone-center modes as indicated in Fig. 4(a). On the other hand, zone-boundary modes tend to be comprised of a mixture of such fundamental polarizations, as the ~ 180meV mode highlighted in Fig. 4(b), which is mainly a combination of longitudinal optical (LO) and transverse acoustic (TA) polarizations. It should also be noted that the frequency of the radial breathing mode (RBM) calculated here is in very good agreement with the relationship derived from *ab initio* calculations,

$$\hbar \omega_{RBM} \approx 28 meV / d_{CNT} \tag{34}$$

where $d_{CNT}$ is the CNT diameter in nanometers [24,25,30].

The Hamiltonian of electron-phonon interaction in a general form is [45]

$$V = \sum_q K_q a_q \left( b_q e^{-i\omega_q t + iqr} + b_q^\dagger e^{i\omega_q t - iqr} \right) \tag{35}$$

where $b_q^\dagger, b_q$ are the creation and annihilation operators for phonons in the mode $q$. The summation over momenta is generally defined via an integral over the first Brillouin zone,

$$\sum_q = \left( \frac{L}{2\pi} \right)^D \int d^D q. \tag{36}$$

where $D$ is the number of unconfined dimensions. For carbon nanotubes $D = 1$ and the limits of the integral are $\pm \pi / (3 a_{cc})$ as follows from (3).

Electron-phonon (e-ph) coupling calculations have also been carried out, as described in [27], in conjunction with the dispersion calculations in order to account for the mode polarization effect on e-ph coupling value [47]. We find that only a few phonon modes that effectively couple to the electrons. As highlighted in Fig. 4(a), out of zone-center modes only the LO (190meV), LA, and radial breathing mode (RBM) have sufficient coupling, whereas, from zone-boundary modes only the 180meV LO/TA mode has significant coupling. Even though we have shown phonon dispersions for a large section of the 1D Brillouin zone, only the ones close to the zone center (i.e.: $q \approx 0$) are involved in electron transport [16]. Within that region of the Brillouin zone all the optical modes are found to have constant energy dispersion while the acoustic mode has a linear dispersion. Thus, in this study all the relevant optical modes for electron transport are considered dispersionless with constant energy, $\hbar\omega_{OP}$, and the zone-center LA mode is taken to be linear with, $\omega_{AP} = v_a q$, relationship where $v_a$ is the sound velocity of that mode. The matrix element of interaction for acoustic phonons is approximated by a linear function $|K_q| = \tilde{K}_a(l)q$. In this paper, we take the matrix elements as inputs and describe the general method of treatment of electron-phonon interaction in nanotubes for both optical and acoustic phonon modes.

## D. Electron-Phonon Scattering

As derived in Appendix B, the in/out-scattering functions for electron-phonon scattering in a ring $i$ from subband $m'$ to subband $m$ are

$$\Sigma_{scat}^{in}(i,i,m,E) = D_0(n_\omega+1)G^n(i,i,m',E+\hbar\omega) + D_0 n_\omega G^n(i,i,m',E-\hbar\omega). \quad (37)$$

$$\Sigma_{scat}^{out}(i,i,m,E) = D_0(n_\omega+1)G^p(i,i,m',E-\hbar\omega) + D_0 n_\omega G^p(i,i,m',E+\hbar\omega). \quad (38)$$

The imaginary part of self-energy is

$$\Sigma_{scat}^i(E) = -\frac{i}{2}\Gamma_{scat}(E) = -\frac{i}{2}\left[\Sigma_{scat}^{in}(E) + \Sigma_{scat}^{out}(E)\right]. \quad (39)$$

The real part of self-energy is manifested as a shift of energy levels and is computed by using the Hilbert transform [37]

$$\Sigma_{scat}^r = P\int \frac{dE'}{2\pi}\frac{\Gamma(E')}{E-E'}. \quad (40)$$

In this paper we neglect the real part of electron-phonon self energy in order to simplify the computations and because the estimates suggest small influence of the real part. For elastic scattering, i.e. in case it is possible to neglect the energy of a phonon, the in/out-scattering energies are

$$\Sigma_{scat}^{in}(i,i,m,E) = D_{el}G^n(i,i,m',E). \quad (41)$$

$$\Sigma_{scat}^{out}(i,i,m,E) = D_{el}G^p(i,i,m',E). \quad (42)$$

In this case there is not need to neglect the real part of self-energy, and its complete expression is

$$\Sigma_{scat}(i,i,m,E) = D_{el}G(i,i,m',E). \quad (43)$$

For optical phonon scattering, the coupling constant is (see Appendix B)

$$D_0 = \frac{\hbar|K_0|^2}{2\rho_{1D}\omega_0\Delta z}. \quad (44)$$

For acoustic phonon scattering, the coupling constant is

$$D_{el} = \frac{\tilde{K}_a^2 k_B T}{\rho_{1D} v_a^2 \Delta z}. \tag{45}$$

In Appendix B, we provide the justification for using only diagonal terms of the self-energy and in/out-scattering functions. We also make the connection (in Appendix C) between the in/out-scattering functions in the coordinate space and the traditionally considered scattering rates in the momentum space.

### III. NUMERICAL TREATMENT OF DISSIPATIVE TRANSPORT

Here, we summarize the overall simulation procedure used in this study. Throughout this work we encounter many energy integrals such as Eqs. (17) and (28). The use of a uniform energy grid becomes prohibitive when sharp features such as quantized energy states need to be accurately resolved. Therefore, an adaptive technique for energy integrations is used based on the *quad.m* subroutine of Matlab® programming language. The treatment of phonon scattering is performed using the self-consistent Born approximation [38, 39]. In that, we need to treat the interdependence of the device Green's function, Eq. (1), and the scattering self-energy, Eq. (2), self-consistently. The treatment of OP scattering is presented first, followed by that for AP scattering.

### *A. Treatment of Optical Phonon Scattering*

The determination of in/out-scattering self-energy functions, Eqs. (3) and (4), for OP scattering requires the knowledge of the electron and hole correlation functions; specifically, the energy-resolved diagonal elements of these functions, $G_{j,j}^{n/p}(E)$. It should be noted that only the diagonal elements are needed since we take the scattering self-energy functions to be diagonal in the local interaction approximation [38, 39]. With that, we use the following procedure to determine $G$ and $\Sigma_{scat}$ self-consistently.

1) Start with known energy-resolved $G_{j,j}^{n/p}$ distributions. Ballistic distributions are used as the starting point.

2) Determine $\Sigma_{scat}^{in}(E)$, $\Sigma_{scat}^{out}(E)$, and $\Sigma_{scat}(E)$ using Eqs. (37), (38), and (39), respectively, at a given energy $E$.

3) Determine new $G(E)$ using Eq. (1).

4) Now, determine new $G^n(E)$ and $G^p(E)$ from Eqs. (11) and (12), respectively.

5) Repeat steps 2 through 4 for all energies and build new $G_{j,j}^{n/p}$ distributions.

6) Repeat steps 1 through 5 until convergence criterion is satisfied. We use the convergence of the induced carrier density, Eq. (28), as the criterion.

In the above calculations, there is a repetitive need for the inversion of a large matrix, Eq. (1), which can be a computationally expensive task. However, we only need a few diagonals of the eventual solution such as the main diagonal of $G^{n/p}$ for the calculation of scattering and carrier densities, and the upper/lower diagonals of $G^n$ for the calculation of current in Eq. (17). The determination of these specific diagonals, in the nearest-neighbor tight-binding scheme, can be performed using the efficient algorithms given in [49]. A Matlab® implementation of these algorithms can be found at [50]. Finally, it should be noted that the overall accuracy of the Born convergence procedure described above is confirmed at the end by observing the current continuity throughout the device, Eq. (17).

*B. Treatment of Acoustic Phonon Scattering*

Similar to the above method, AP scattering is treated using the following procedure,

1) Start with known energy-resolved $G_{j,j}^{n/p}$ distributions. Ballistic distributions are used as the starting point.

2) Determine $\Sigma_{scat}^{in}(E)$, $\Sigma_{scat}^{out}(E)$, and $\Sigma_{scat}(E)$ using Eqs. (41), (42), and (43), respectively, at a given energy $E$.

3) Determine new $G(E)$ using Eq. (1).

4) Now, determine new $G^n(E)$ and $G^p(E)$ at energy $E$ from Eqs. (11) and (12), respectively.

5) Repeat steps 2 through 4 until convergence criterion is satisfied. Here, we use the convergence of $G^n(E)$.

6) Repeat steps 2 thru 5 for all energies and build new $G_{j,j}^{n/p}$ distributions.

7) Repeat steps 1 through 6 until convergence criterion is satisfied. We use the convergence of the induced carrier density, Eq. (28), as the criterion.

For the case of AP scattering we have introduced an additional convergence loop (step 5 above) since, unlike in inelastic scattering, here the self-consistent Born calculation at a given energy is decoupled from that at all other energy values. Similar to OP scattering, we use the efficient algorithms of [49] for numerical calculations, and confirm the overall accuracy of the convergence procedure by monitoring current continuity throughout the device.

## IV. RESULTS AND DISCUSSION

Dissipative transport simulations are carried out as explained in the previous sections, and the results are compared to that with ballistic transport. Here, we first study the effects of phonon scattering on CNTFET characteristics using a (16,0) tube as a representative case. Then, we compare the diameter dependence using (16,0), (19,0) and (22,0) tubes, that belong to the *mod(n-m,3)* = 1 family. The device parameters (Fig. 1(a)) used for the simulation of OP scattering are as follows: $L_{ch}$ = 20nm, $L_{SD}$ = 30nm, $L_{ext}$ = 0, $t_{ox}$ = 2nm (HfO$_2$ with $\kappa$ = 16), and the source/drain doping $N_{SD}$ = 1.5/nm. This doping concentration should be compared with the carbon atom density of ($4n/3a_{cc}$) in an (*n,0*) zigzag CNT, which is ~ 150/nm in a (16,0) tube. For the simulation of AP scattering, a heavy doped extension region is used for better convergence of the electrostatic solution. In this case, $L_{SD}$ = 20nm, $L_{ext}$ = 15nm, $N_{SD}$ = 1.5/nm, and the extension doping, $N_{ext}$ = 1.8/nm are used; and all the other parameters are same as for the previous case. Except for assisting in the convergence procedure, the effect of the heavy doped extensions on the device characteristics is negligible. It should be noted that under OP scattering we consider the impact of intra-LO, intra-RBM, and inter-LO/TA phonon modes all together simultaneously (Table I). The intra-LA mode is treated under AP scattering separately.

Figure 5 compares the $I_{DS}$-$V_{DS}$ results for the (16,0) CNTFET under ballistic transport and that with OP and AP scattering. It is seen that phonon scattering can indeed have an appreciable effect on the device ON-current: at $V_{GS}$ = 0.6V the ON-current is reduced by ~ 9% and ~ 7% due to OP and AP scattering, respectively. The relative importance of the two scattering mechanisms also shows an interesting behavior. Up to moderate gate biases the effect of AP scattering is stronger ($V_{GS} \leq$ 0.5V). At large gate biases OP scattering becomes the more important process ($V_{GS} \geq 0.6V$). This relative behavior can be better observed in the $I_{DS}$-$V_{GS}$ results shown in Fig. 6. Here, it is seen that up to moderate gate biases AP scattering causes a larger reduction in the device current compared to OP scattering. Even in this case, the current reduction seen for OP scattering is mainly due to the low-energy RBM mode [32]. At large gate biases, however, the effect of OP scattering becomes stronger, reducing the current by ~ 16% from the ballistic level at $V_{GS}$ = 0.7V. Previous studies have shown that the strong current degradation at larger gate biases is due to high-energy OP scattering processes becoming effective (mainly inter-LO/TA and intra-LO modes) [31,32]. Nevertheless, the importance of AP and low-energy RBM scattering should be appreciated since these might be the relevant scattering mechanisms under typical biasing conditions of a nanoscale transistor.

The relative behavior of OP and AP scattering can be understood by studying Fig. 7. It shows the energy-position resolved current spectrum, which is essentially the integrand of Eq. (17), under ballistic transport and OP scattering. In Fig. 7(a), it is seen that under ballistic conditions, carriers injected from the source reaches the drain without losing energy inside the device region. There exists a finite density of current below the conduction band edge ($E_C$) which is due to quantum mechanical tunneling. In the presence of OP scattering, however, it is seen that the carriers near the drain end relaxes to low energy states by emitting phonons (Fig. 7(b)). Nevertheless, up to moderate gate biases high-energy OP scattering does not affect the device current due to the following reason. For such biasing conditions the energy difference between the source Fermi level and the top of the channel barrier, $\eta_{FS}$, is smaller than the optical phonon energy: $\eta_{FS} \ll \hbar\omega_{OP}$. Therefore, a majority of the positive going carriers (source → drain) in the channel region does not experience high-energy OP scattering, except for a minute portion in the high-energy tail of the source Fermi distribution. On the other hand, when these carriers reach the drain end there are empty low-lying states that they scatter to. After emitting a high-energy OP, however, these carriers do not have enough energy to surmount the channel barrier and reach the source region again. Thus, the effect of high-energy OP scattering on the device current is suppressed until backscattering becomes effective at larger gate biases for $\eta_{FS} \geq \hbar\omega_{OP}$. On the other hand, low-energy RBM phonons and acoustic phonons can effectively backscatter at all gate biases. They are the dominant scattering mechanism until high-energy OP becomes important at large biases [31,32].

Figure 8 shows the energy-position resolved electron density spectrum, which is essentially the integrand of Eq. (14). Examining Fig. 8(a), one can see that electrons are filled up to the respective Fermi levels in the two contact regions. In these regions, a characteristic interference pattern in the distribution function is observed due to quantum mechanical inference of positive and negative going states [42]. Quantized valence band states in the channel region are due to the longitudinal confinement in this effective potential well [42]. In the presence of OP scattering, few interesting features are observed in Fig. 8(b). The interference pattern seen in the contact regions are smeared due to the broadening of energy states by incoherent OP scattering. The electrons near the drain end relax down to low lying empty states, even though they are less discernible in the linear color scale employed here. More interestingly, now we observe a multitude of quantized valence band states in the channel region. Such states with energies below the conduction band edge of the drain region are observed here due to their additional broadening by coupling to the phonon bath. They were unobservable in the ballistic case since they lied inside the bandgap regions of the contact reservoirs that led to zero contact broadening, $\Gamma_{S/D} \approx 0$. The additional low-intensity states observed are the phonon induced side-bands of the main quantized levels originating from the variety of OP modes considered here. Carrier transport through these quantized states is indeed possible under appropriate biasing conditions, and lead to many interesting properties such as, less than 60mV/decade subthreshold operation and, phonon-assisted inelastic tunneling. The interested reader is referred to [34,35].

Figure 9 explores the diameter dependence of the impact of phonon scattering in CNTFETs. As mentioned earlier, we consider the *mod(n-m,3) = 1* type of tubes. Similar trends in the behavior can be expected for the *mod(n-m,3) = 2* family as well [28,29]. Here we compare the ballisticity of tubes, defined as the ratio between current under scattering and the ballistic current ($I_{scat}/I_{ballist}$), vs. $\eta_{FS}$, defined in Fig. 7(b). Positive $\eta_{FS}$ corresponds to the on-state of the device at large positive gate biases, and negative $\eta_{FS}$ is for the off-state. The characteristic roll-off of ballisticity under OP scattering is seen in Fig. 9(a) [32]. In that, the roll-off is due to high-energy OP scattering becoming effective at large gate biases. The ballisticity reduction at small gate biases is due to the low-energy RBM scattering [32]. In Fig. 9(a) it is seen that the impact of high-energy OP scattering decreases for larger diameter tubes. This can be easily understood by noting that the e-ph coupling parameter for these modes (intra-LO and inter-LO/TA) monotonically decreases with increasing diameter (Table I). On the other hand, the impact of the RBM mode at low gate biases seems to be nearly diameter independent for the tubes considered here, even though there is a similar decrease in e-ph coupling for larger diameter tubes (Table I). This behavior is due to the concomitant reduction of energy of the RBM mode at larger diameters that leads to an increased amount of scattering events, which ultimately cancels out the overall impact on device current.

Diameter dependence of AP scattering is shown in Fig. 9(b). The ballisticity for larger tubes is higher due to the corresponding reduction of the e-ph coupling parameter shown in Table I. They all show a slight increase in the ballisticity at larger gate biases due to majority of the positive going carriers occupying states well above the channel conduction band edge [32]. The backscattering rate is a maximum near the band edge due to increased 1D density of states and decays at larger energies [14,16,18]. It is seen that for all the tubes on Fig. 9, the impact of AP scattering is stronger compared to OP scattering until the high-energy modes become effective. Under typical biasing conditions for nanoscale transistor operation, $\eta_{FS}$ will be limited ($\eta_{FS} \leq 0.15$eV) and the transport will be dominated by AP and low-energy RBM scattering [51].

## V. CONCLUSION

In conclusion, we present here the detailed self-contained description of the NEGF method to simulate transport of carriers in carbon nanotube transistors with the account of both quantum effects and electron-phonon scattering. This capability is especially necessary, since it provides the rigorous treatment in the practically important limit of intermediate length devices. We outline our numerical procedure for solution of the NEGF equations via convergence of several self-consistent loops. Finally we display a few of the simulation results obtained by this method, such as the energy spectra of carrier density and current, and, current-voltage characteristics. They enable a researcher to uncover the workings of the quantum phenomena underlying the operation of carbon nanotube transistors, and to predict their performance.

We acknowledge the support of this work by the NASA Institute for Nanoelectronics and Computing (NASA INAC NCC 2-1363), NASA contract NAS2-03144 to UARC, and Intel Corporation. Computational support was provided by the NSF Network for Computational Nanotechnology (NCN). S.O.K thanks the Intel Foundation for PhD Fellowship support.

**Appendix A. Notation conventions for Green's functions**

For the benefit of the reader we provide the conversion formulas between the two widely used notation conventions in the NEGF method. The one used in this paper is more intuitive for the device application and is based on Datta's book [37]. Another is traditional in condensed matter physics and is exemplified by [38]. These equivalent notations are shown on the left and right, respectively

$$G \leftrightarrow G^r, \tag{46}$$

$$G^\dagger \leftrightarrow G^a, \tag{47}$$

$$G^n \leftrightarrow -iG^<, \tag{48}$$

$$G^p \leftrightarrow iG^>, \tag{49}$$

$$\Sigma \leftrightarrow \Sigma^r, \tag{50}$$

$$\Sigma^\dagger \leftrightarrow \Sigma^a, \tag{51}$$

$$\Sigma^{in} \leftrightarrow -i\Sigma^<, \tag{52}$$

$$\Sigma^{out} \leftrightarrow i\Sigma^>. \tag{53}$$

**Appendix B. Derivation of the in/out-scattering energies for electron-phonon interaction.**

Though the self-energy for the electron-phonon scattering has been discussed multiple times, e.g. [37], considerable confusion still exists about its form and assumptions used in the derivation. One reason may be the fact that in device simulation one uses Green's functions and self-energy functions of two coordinate arguments, while the scattering processes are traditionally formulated in the momentum-dependent and coordinate independent representation. The other reason is that the expression for self energy looks slightly different for different material systems. Here we aim to derive the expression for the self-energy in a simple, but general form, and then to specify it for the particular case of one-dimensional transport in nanotubes.

The self consistent Born approximation results in the following in- and out-scattering functions for the electron-phonon interactions [38,52]

$$\Sigma^{in,out}(X_1, X_2) = G^{n,p}(X_1, X_2) D^{n,p}(X_1, X_2). \tag{54}$$

where the argument $X = \{r, m, t\}$ incorporates the spatial coordinates in the unconfined dimensions, subband/valley index, and time, respectively. The phonon propagator contains the average over the random variables of the reservoir designated by angle brackets

$$D^n(X_1, X_2) = \langle V(X_1)V(X_2) \rangle, \quad D^p(X_1, X_2) = \langle V(X_2)V(X_1) \rangle \tag{55}$$

The averages of the following operator products in a reservoir at thermal equilibrium depend on the phonon occupation numbers (33)

$$\langle b_q^\dagger b_{q'} \rangle = \delta_{qq'} n_q, \quad \langle b_q b_{q'}^\dagger \rangle = \delta_{qq'}(n_q + 1), \tag{56}$$

and all other averages of pair products are zero. On substitution of the electron-phonon Hamiltonian (35) it results in

$$D^n(r_1,m_1,t_1,r_2,m_2,t_2) = \sum_q |K_q|^2 a_q^2$$
$$\left[(n_q+1)\exp(i\omega_q(t_2-t_1)+iq(r_1-r_2)) + n_q \exp(i\omega_q(t_1-t_2)+iq(r_2-r_1))\right]$$

(57)

and a similar expression for $D^p(r_1,l_1,t_1,r_2,l_2,t_2)$. The selection rules for the electron subbands, $m,m'$, and phonon subbands, $l$, is as described in Section II.C. Then we limit the consideration to stationary situation, i.e. where the functions depend only on the difference of times $t = t_2 - t_1$. The Fourier transform relative to this time interval produces energy-dependent in/out-scattering functions (given here for a specific phonon subband)

$$\Sigma^{in}(r_1,r_2,m,E) = D(r_1,r_2,l,E)(n_q+1)G^n(r_1,r_2,m',E+\hbar\omega_q)$$
$$+ D^*(r_1,r_2,l,E)n_q G^n(r_1,r_2,m',E-\hbar\omega_q)$$

(58)

$$\Sigma^{out}(r_1,r_2,m,E) = D^*(r_1,r_2,l,E)(n_q+1)G^p(r_1,r_2,m',E-\hbar\omega_q)$$
$$+ D(r_1,r_2,l,E)n_q G^p(r_1,r_2,m',E+\hbar\omega_q)$$

(59)

where the first term in the expressions corresponds to emission of a phonon, and the second one – to absorption of a phonon. The electron-phonon coupling operator contains the sum over the phonon momentum that operates on the factors to the right of it

$$D(r_1,r_2,l) = \sum_q |K_q|^2 a_q^2 \exp(-iqr).$$

(60)

It depends on the difference of the spatial coordinates $r = r_2 - r_1$. The expressions for the in/out-scattering functions drastically simplify in the two following cases.

First, for isotropic scattering with phonons of constant energy ($|K_q| \approx |K_0|$ and $\omega_q \approx \omega_0$, and they are independent of $q$). This is approximately fulfilled for optical phonons. In this case, the electron-phonon scattering operator reduces to calculation of a sum

$$D(r_1, r_2, l) = \frac{\hbar |K_0|^2}{2\rho_{1D}\omega_0} \int_{-\pi/(3a_{cc})}^{-\pi/(3a_{cc})} \frac{dq}{2\pi} \exp(-iqr). \qquad (61)$$

For the distance of integer multiple of the nanotube period $r = j3a_{cc}$, the integral above

$$\int_{-\pi/(3a_{cc})}^{-\pi/(3a_{cc})} \frac{dq}{2\pi} \exp(-iqr) = \begin{cases} 1/(3a_{cc}), j = 0 \\ 0, j \neq 0 \end{cases}. \qquad (62)$$

One needs to insert the factor of 4, for the number of rings in the period, to obtain that the electron-phonon coupling is a constant factor (44)

$$D_0 = \frac{\hbar |K_0|^2}{2\rho_{1D}\omega_0 \Delta z}. \qquad (63)$$

and the expression for the in/out-scattering functions (37) and (38). Also, a very important conclusion is that the self-energy and the in/out-scattering functions can be treated as diagonal in this case. This significantly simplifies the problem and permits the use of various algorithms of solution of the matrix equations only applicable to 3-diagonal matrices, such as the recursive inversion method [38].

Second case, for elastic scattering, when one can neglect the energy of a phonon compared to characteristic energy differences. This is approximately fulfilled for acoustic phonons. For this case, the dependence on the momentum is typically $\omega_q = v_a(l)q$ and $|K_q| = \tilde{K}_a(l)q$; and only phonons with momentum close to $q = 0$ have the appreciable occupations, such that

$$n_q \approx \frac{k_B T}{\hbar \omega_q} \gg 1. \tag{64}$$

Then again as in (61), the matrix element and the number of phonon factors prove to be independent of the phonon momentum and can be taken out of the summation

$$\Sigma^{in}(r_1, r_2, m, E) = G^n(r_1, r_2, m', E) \int_{-\pi/(3a_{cc})}^{-\pi/(3a_{cc})} \frac{k_B T}{\hbar \omega_q} \frac{\hbar \tilde{K}_a^2 q^2}{2\rho_{1D} \omega_q} \frac{dq}{2\pi} \exp(-iqr) + c.c. \tag{65}$$

to again yield a diagonal in/out-scattering functions (41) and (42)

$$\Sigma^{in}(r_1, r_1, m, E) = 2 \frac{k_B T \tilde{K}_a^2}{2\rho_{1D} v_a^2} G^n(r_1, r_1, m', E) \frac{4}{3a_{cc}} \tag{66}$$

and the constant elastic electron-phonon coupling (45). Note an additional factor of 2 in these expressions because the processes with emission and absorption of a phonon are now lumped into one term.

By going beyond the assumption of a constant product of the coupling factor and the phonon occupation, we can determine how good the approximation of a diagonal self-energy is. By representing it as a Taylor series (and we know that it is an even function)

$$|K_q|^2 a_q^2 n_q = |K_0|^2 a_0^2 n_0 \left(1 + \frac{q^2}{q_{(2)}^2} + ...\right). \tag{67}$$

and examining the second term, we obtain

$$\int_{-\pi/(3a_{cc})}^{-\pi/(3a_{cc})} \frac{dq}{2\pi} \cdot \frac{q^2}{q_{(2)}^2} \exp(-iqr) = \begin{cases} \pi^2/(q_{(2)}^2 3^4 a_{cc}^3), j = 0 \\ 2(-1)^j/(jq_{(2)}^2 3^3 a_{cc}^3), j \neq 0 \end{cases}. \tag{68}$$

This can be restated as: the off-diagonal terms of the self energy and the in/out-scattering functions have the order of magnitude of the variation of the product (67) over the first

Brillouin zone. By doing an inverse Fourier transform of (67), we recognize the parameter $q_{(2)}$ as the inverse characteristic radius of electron-phonon interaction. Thus the alternative formulation of the above criterion is: the self-energy is diagonal if the corresponding interaction radius is much less than the crystal lattice size.

**Appendix C. Connection between self-energy, scattering rates, and mean free path.**

In this section we draw the correspondence between the in/out-scattering functions and the scattering rates, which researchers typically deal with in the classical description of transport. The probability of scattering between two specific momentum states $p$ and $p'$ of carriers is calculated according to Fermi's "golden rule" [45]

$$S(p,p') = \frac{2\pi}{\hbar}|K_q|^2 a_q^2 \left(n_q + \frac{1}{2} \mp \frac{1}{2}\right) \delta_{p',p\pm q} \delta(E' - E \mp \hbar\omega_q). \tag{69}$$

where the upper sign corresponds to absorption of a phonon and the lower sign – to emission of a phonon. The total scattering rate for carriers with momentum $p$ is

$$\frac{1}{\tau(p)} = \sum_{p'} S(p,p'). \tag{70}$$

where summation is performed only over momentum variables but not the spin variables. In other words, the spin state is assumed unchanged in scattering. For isotropic scattering, such as deformation potential of acoustic of optical phonons, the scattering rate (70) is equal to the momentum relaxation rate [45]. Also in this case, the momentum summation can be replaced with the help of Eq. (36) by the integral over energies

$$\sum_{p'} = L\int \frac{d^D k}{(2\pi)^D} = \frac{L}{2}\int_0^\infty dE\, g_D(E). \tag{71}$$

this yields

$$\frac{1}{\tau(E)} = \frac{\pi L}{\hbar}|K_q|^2 a_q^2 \left(n_q + \frac{1}{2} \mp \frac{1}{2}\right) g_D(E \pm \hbar\omega_q). \tag{72}$$

A general expression for the scattering rate (for one-dimensional structures) is

$$\frac{1}{\tau(E)} = \frac{2\pi}{\hbar} R_0 \cdot \left(n_q + \frac{1}{2} \mp \frac{1}{2}\right) g_{1D}(E \pm \hbar\omega_q) \tag{73}$$

For electron-phonon scattering, the constant in this expression is related to the constant in the in/out-scattering functions as

$$R_0 = \frac{\hbar |K_q|^2}{4\rho_D \omega_q} = \frac{D_0 \Delta z}{2}. \tag{74}$$

Similarly one obtains for elastic scattering (both with emission and absorption of phonons)

$$\frac{1}{\tau_{el}(E)} = \frac{2\pi}{\hbar} R_{el} g_{1D}(E) \tag{75}$$

with a similar relation between the constant in the scattering rate and in the in/out-scattering functions

$$R_{el} = \frac{\tilde{K}_a^2 k_B T}{2\rho_{1D} v_a^2} = \frac{D_{el} \Delta z}{2}. \tag{76}$$

Consider for example an in-scattering function with phonon emission. It must be equal to the rate of in-coming particles multiplied by the Planck's constant.

$$\Sigma^{in,em}(E) = \frac{\hbar}{\tau(E)} n_{el}(E + \hbar \omega_q) = \frac{\hbar}{\tau(E)} \frac{G^n(E + \hbar \omega_q)}{A(E + \hbar \omega_q)}. \tag{77}$$

With the help of Eqs. (16) and (73) it reduces to

$$\Sigma^{in,em}(E) = 2R_0 \cdot (n_q + 1) \frac{G^n(E + \hbar \omega_q)}{\Delta z}. \tag{78}$$

which does, in fact, coincide with the first term in (37).

The mean free path for carriers of certain energy is given by the product their velocity and scattering time

$$\lambda(E) = v(E)\tau(E). \tag{79}$$

By substituting the scattering rate (73) and the density of states (25), we obtain for the mean free path relative to scattering with emission of a phonon as

$$\lambda(E) = \frac{\hbar^2 v(E) v(E + \hbar \omega_q)}{4 R_0 \cdot (n_q + 1)}. \tag{80}$$

This expression simplifies in the limit of high enough energies, i.e., far from the band edge, according to (26). We also take the limit of phonon occupation number $n_q \ll 1$

$$\lambda_{hi} = \frac{\hbar^2 v_F^2}{4 R_0} = \frac{9 t^2 a_{cc}^2}{16 R_0}. \tag{81}$$

Not that the same form of equation is valid for elastic scattering, though with $n_q \gg 1$. Recalling the in/out-scattering function constant (74)

$$D_0 = \frac{3 t^2 a_{cc}}{2 \lambda_{hi}}. \tag{82}$$

The above nominal mean free path (81) is the upper limit over all energies. In semiconducting nanotubes, velocity is smaller for energies closer to the band edge, and the density of states is larger. Therefore the specific mean free path is shorter for energies closer the band edge, and likewise, the mean free path averaged over the carriers' distribution can be orders of magnitude shorter than (81). Therefore scattering can be significant in a 20nm-channel transistor even if the nominal mean free path is close to 1 micrometer. However the value for the nominal mean free path is sometimes used as a parameter in experiments. Note that it would provide a good estimate for the mean free path in metallic nanotubes, which have zero band gap and linear energy dispersion.

**List of Table Captions**

**TABLE 1. Phonon energy and e-ph coupling parameters for the CNTs used in this study.**

**List of Tables**

| Phonon mode | (16,0) <br> d = 1.25nm, $E_G$ =0.67eV | (19,0) <br> d = 1.50nm, $E_G$ =0.56eV | (22,0) <br> d = 1.70nm, $E_G$ =0.49eV |
|---|---|---|---|
| Intra LO (190meV)[a] | 9.80x10$^{-3}$ eV$^2$ | 8.19x10$^{-3}$ eV$^2$ | 7.00x10$^{-3}$ eV$^2$ |
| Intra RBM[a,b] | 0.54x10$^{-3}$ eV$^2$ (21meV) | 0.36x10$^{-3}$ eV$^2$ (18meV) | 0.25x10$^{-3}$ eV$^2$ (16meV) |
| Inter LO/TA (180meV)[a] | 19.30x10$^{-3}$ eV$^2$ | 16.26x10$^{-3}$ eV$^2$ | 14.13x10$^{-3}$ eV$^2$ |
| Intra LA[c] | 2.38x10$^{-3}$ eV$^2$ | 2.00x10$^{-3}$ eV$^2$ | 1.73x10$^{-3}$ eV$^2$ |

**TABLE 1. Phonon energy and e-ph coupling parameters for the CNTs used in this study.**

a) e-ph coupling for optical phonons is determined according to Eq. (44);

b) RBM energy is diameter dependent, and shown in the parentheses;

c) e-ph coupling for acoustic phonons is determined according to Eq. (45).

**List of Figure Captions**

Fig. 1. (color online) (a) Device structure with wrap-around gate, (b) NEGF model with coupling to the phonon bath, and (c) mode-space Hamiltonian.

Fig. 2. (color online) Lowest energy degenerate subbands in a CNT corresponding to K and $K^/$ valleys of 2D graphene Brillouin zone. (a) and (b) show intra-valley and inter-valley scattering processes, respectively.

Fig 3. (color online) Self-consistency requirement between NEGF and Poisson solutions.

Fig 4. (color online) Energy dispersion for phonon modes in a (16,0) CNT: (a) zone-center phonons that allow intra-valley scattering and, (b) zone-boundary phonons that allow inter-valley scattering. Modes that effectively couple to the electrons are indicated by dashed circles. Zone-boundary phonons are composed of a mixture of fundamental polarizations.

Fig 5. (color online) $I_{DS}$-$V_{DS}$ for the (16,0) CNTFET under ballistic transport, OP scattering (all modes together), and AP scattering. High-energy OP scattering becomes important at sufficiently large gate biases. Until then AP and RBM scattering are dominant.

Fig 6. (color online) $I_{DS}$-$V_{GS}$ for the (16,0) CNTFET at $V_{DS}$ = 0.3V under ballistic transport, OP scattering (all modes together), and AP scattering. The inset shows that acoustic phonons are more detrimental up to moderate gate biases.

Fig. 7. (color online) Energy-position resolved current spectrum for (16,0) CNTFET at $V_{GS}$ = 0.5V, $V_{DS}$ = 0.5V (logarithmic scale). (a) ballistic, (b) dissipative transport (all OP modes together). Thermalization near the drain end by emitting high-energy OPs leaves the electrons without enough energy to overcome the channel barrier.

Fig. 8. (color online) Energy-position resolved electron density spectrum for (16,0) CNTFET at $V_{GS}$ = 0.5V, $V_{DS}$ = 0.5V. (a) ballistic, (b) dissipative transport (all OP modes together). Quantized states in the valence band are broadened, and give rise to many phonon induced side-bands. The interference pattern for conduction band states are also broadened compared to the ballistic case.

Fig. 9. (color online) Ballisticity ($I_{scat}/I_{ballist}$) vs. $\eta_{FS}$ for (16,0), (19,0) and (22,0) CNTFETs, (a) with all OP modes together, (b) with AP scattering. $\eta_{FS}$ is defined as the energy difference between the source Fermi level and the channel barrier (see Fig. 8(b)).

**List of Figures**

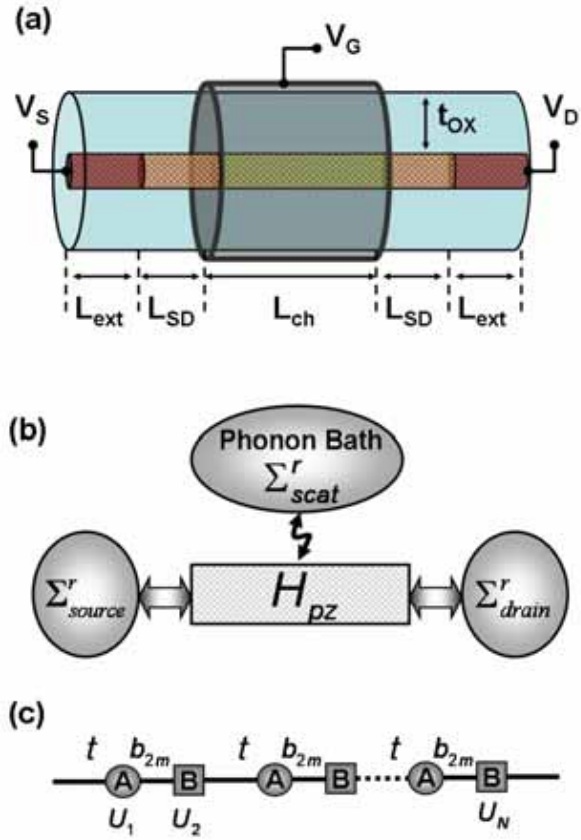

Fig. 1. (color online) (a) Device structure with wrap-around gate, (b) NEGF model with coupling to the phonon bath, and (c) mode-space Hamiltonian.

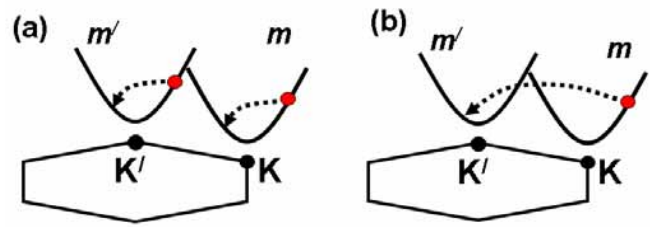

Fig. 2. (color online) Lowest energy degenerate subbands in a CNT corresponding to K and K$^{/}$ valleys of 2D graphene Brillouin zone. (a) and (b) show intra-valley and inter-valley scattering processes, respectively.

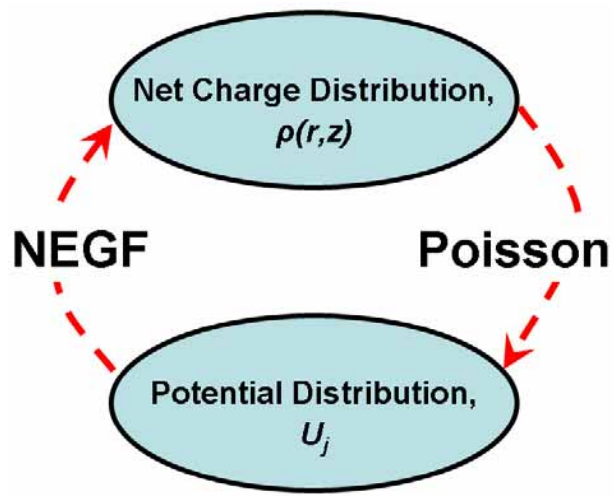

Fig 3. (color online) Self-consistency requirement between NEGF and Poisson solutions.

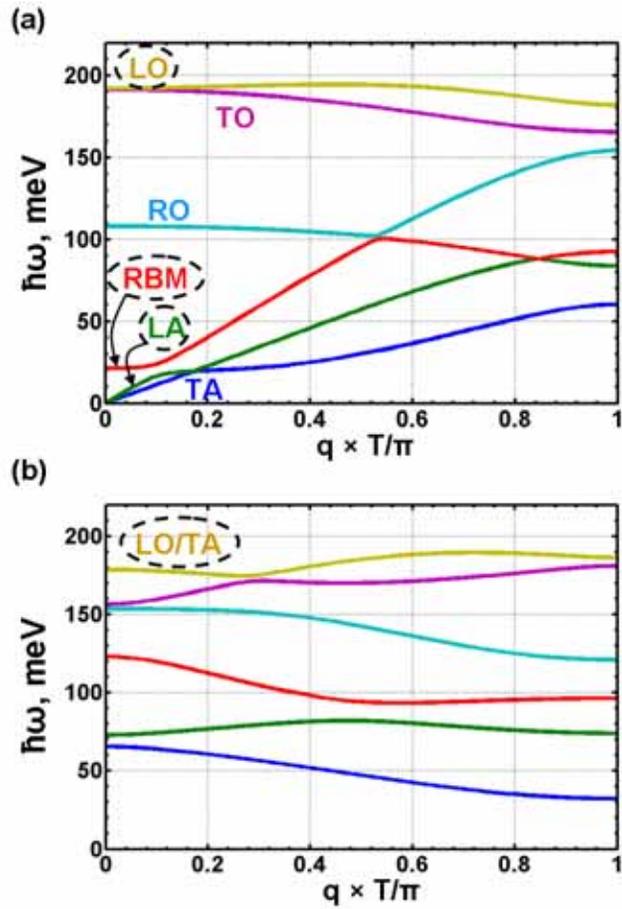

Fig 4. (color online) Energy dispersion for phonon modes in a (16,0) CNT: (a) zone-center phonons that allow intra-valley scattering and, (b) zone-boundary phonons that allow inter-valley scattering. Modes that effectively couple to the electrons are indicated by dashed circles. Zone-boundary phonons are composed of a mixture of fundamental polarizations.

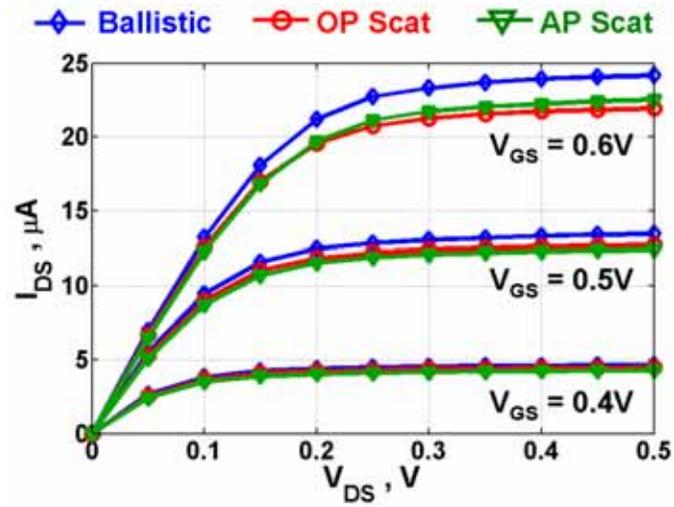

Fig 5. (color online) $I_{DS}$-$V_{DS}$ for the (16,0) CNTFET under ballistic transport, OP scattering (all modes together), and AP scattering. High-energy OP scattering becomes important at sufficiently large gate biases. Until then AP and RBM scattering are dominant.

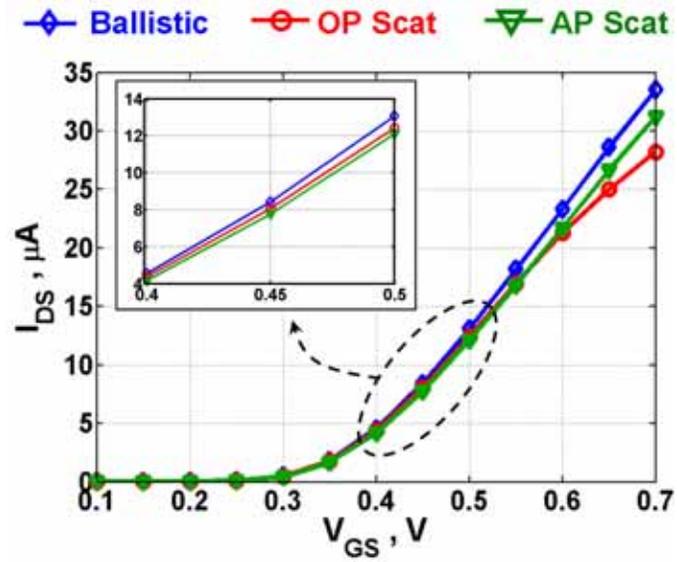

Fig 6. (color online) $I_{DS}$-$V_{GS}$ for the (16,0) CNTFET at $V_{DS}$ = 0.3V under ballistic transport, OP scattering (all modes together), and AP scattering. The inset shows that acoustic phonons are more detrimental up to moderate gate biases.

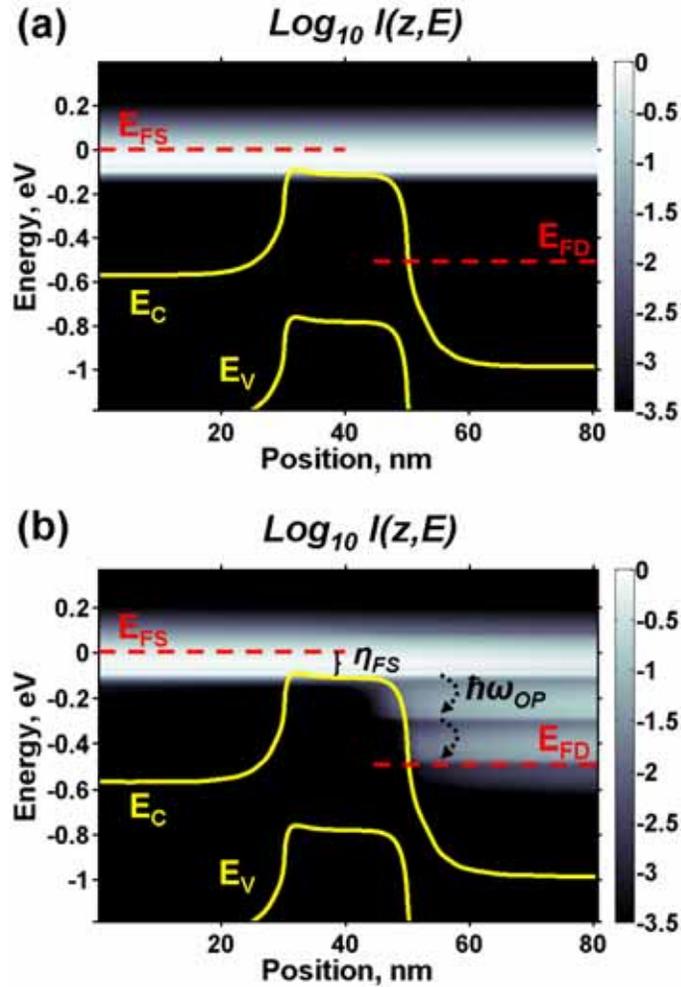

Fig. 7. (color online) Energy-position resolved current spectrum for (16,0) CNTFET at $V_{GS}$ = 0.5V, $V_{DS}$ = 0.5V (logarithmic scale). (a) ballistic, (b) dissipative transport (all OP modes together). Thermalization near the drain end by emitting high-energy OPs leaves the electrons without enough energy to overcome the channel barrier.

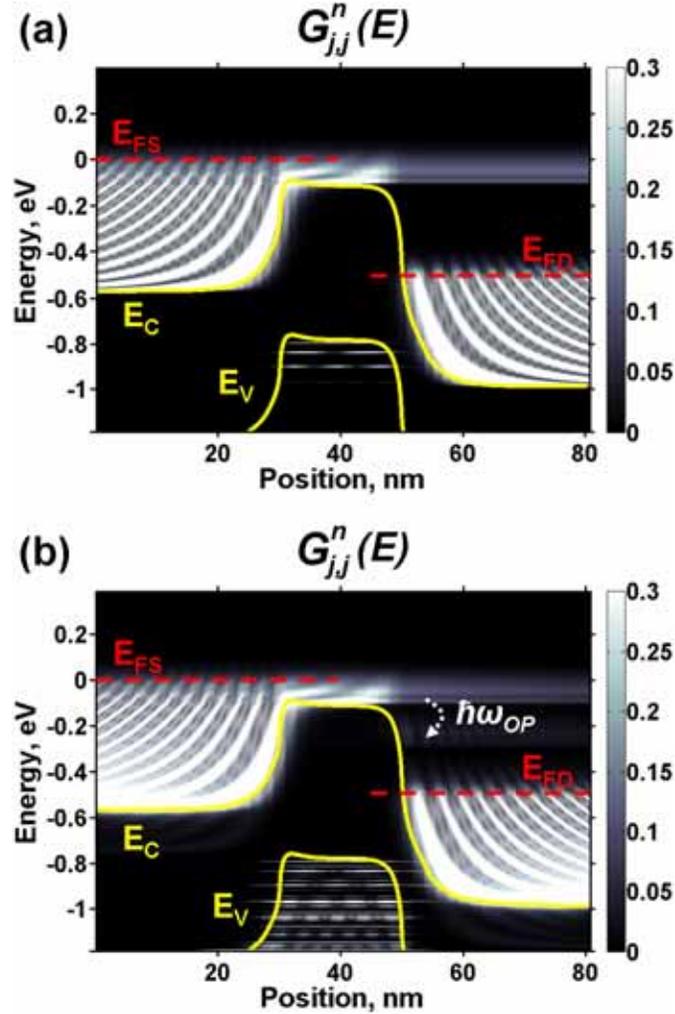

Fig. 8. (color online) Energy-position resolved electron density spectrum for (16,0) CNTFET at $V_{GS}$ = 0.5V, $V_{DS}$ = 0.5V. (a) ballistic, (b) dissipative transport (all OP modes together). Quantized states in the valence band are broadened, and give rise to many phonon induced side-bands. The interference pattern for conduction band states are also broadened compared to the ballistic case.

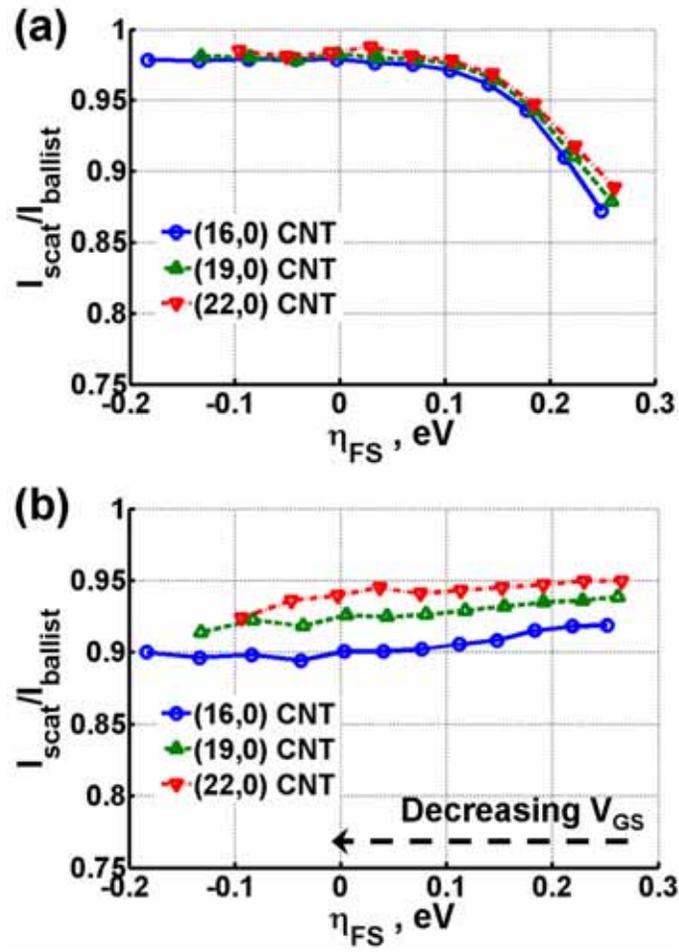

Fig. 9. (color online) Ballisticity ($I_{scat}/I_{ballist}$) vs. $\eta_{FS}$ for (16,0), (19,0) and (22,0) CNTFETs, (a) with all OP modes together, (b) with AP scattering. $\eta_{FS}$ is defined as the energy difference between the source Fermi level and the channel barrier (see Fig. 8(b)).